\documentclass{aa}  

\usepackage{natbib} 
\usepackage{graphicx}
\usepackage{xcolor}
\usepackage{txfonts}
\usepackage{amsmath}
\usepackage{mathtools}
 
\usepackage{paralist}

\usepackage{float}
\usepackage{booktabs}
\usepackage{multirow}
\usepackage{booktabs}

\usepackage{multicol}
\usepackage{lipsum, babel}

\begin{document}

   \title{A generalized dipole-segment model for the gravitational field of elongated bodies}

   \subtitle{}

   \author{A. K. de Almeida Jr.
          \inst{1,2}
          \and
          A. F. S. Ferreira\inst{3$\dag$}
          \and
          L. B. T. Santos\inst{4}\fnmsep\thanks{corresponding author}
          \and
          F. Monteiro\inst{3}
          \and
          A. Amarante\inst{3}
          \and
          E. Tresaco\inst{5}
          \and
          Sanchez, D. M.\inst{6}
          \and
          C. Gomes \inst{3}
          \and
          Prado, A. F. B. A.\inst{7}
          }

   \institute{CFisUC, Departamento de Física, Universidade de Coimbra, 3004-516 Coimbra, Portugal \and 
             CICGE, DGAOT, FCUP, Vila Nova De Gaia, Portugal\\
             \email{allan.junior@ua.pt}
         \and
             Department of Mathematics, School of Engineering and Sciences - São Paulo State University (FEG/UNESP), Av. Ariberto Pereira da Cunha 333, 12516-410 Guaratinguetá, Brazil\\
             $\dag$\email{alessandra.ferraz@unesp.br}
         \and
             PostGrad Program in Systems Engineering (PPGES) - University of Pernambuco (UPE), Brazil\\
             $\star$\email{leonardo.torres@upe.br}
        \and
             Instituto Universitario de Matemáticas y Aplicaciones - Universidad de Zaragoza. EPS, Crta. de Cuarte s/n 22071, Huesca, Spain.
             \and
             School of Aerospace and Mechanical Engineering
The University of Oklahoma
865 Asp Ave., 73019 OK, USA
             \and
             National Institute for Space Research (INPE), Av. dos Astronautas 1.758, 12227-010 São José dos Campos, SP, Brazil\\
}

%   \date{Received September 15, 1996; accepted March 16, 1997}

% \abstract{}{}{}{}{} 
% 5 {} token are mandatory
 
  \abstract
  % context heading (optional)
  % {} leave it empty if necessary  
   {Various simplified models have been investigated to understand the complex dynamical environment near irregular asteroids.}
  % aims heading (mandatory)
   {We propose a generalized dipole-segment model (GDSM) to describe the gravitational fields of elongated bodies. The proposed model extends the dipole-segment model (DSM) by including variable pole masses and a connecting rod while also accounting for the spheroidal shape of the poles instead of assuming point masses.}
  % methods heading (mandatory)
   {A nonlinear optimization method was employed to determine the model parameters, which minimizes the errors between the equilibrium points predicted by the GDSM and those obtained using a more realistic approach, such as the polyhedron model, which is assumed to provide the accurate values of the system. The model was applied to three real irregular bodies: the Kuiper belt objects Arrokoth, Kleopatra, and comet 103P/Hartley.}
  % results heading (mandatory)
   {The results show that the GDSM represents the gravitational field more accurately than the DSM and significantly reduces computational time and effort when compared with the polyhedron model. This reduction in computational complexity does not come at the cost of efficiency. This makes the GDSM a valuable tool for practical applications. The model was further employed to compute heteroclinic orbits that connect the unstable triangular equilibrium points of the system. These trajectories, obtained from the intersections of the stable and unstable manifolds, represent natural pathways that enable transfers between equilibrium regions without continuous propulsion. The results for Arrokoth, Kleopatra, and 103P/Hartley are consistent and validate the GDSM as an accurate and computationally efficient framework for studying the dynamical environment and transfer mechanisms around irregular small bodies.
}
  % conclusions heading (optional), leave it empty if necessary 
   {}


   \maketitle
%
%-------------------------------------------------------------------

\section{Introduction}

Over the past few decades, space agencies have increasingly directed their efforts toward exploring the small bodies that populate our Solar System by dispatching spacecraft to study asteroids and comets, with further missions planned for the future.
The shapes of asteroids and comets are often irregular. This poses challenges in gravitational modeling. When space missions are designed to explore these celestial bodies, a critical question is how the gravitational field of these small irregularly shaped objects can be represented accurately \citep{Elipe1, 2003CeMDA..86..131B}. To address this challenge, numerous mathematical models have been proposed that each offer potential solutions to better understand and characterize the gravitational effect of these bodies.

Spherical harmonic expansion is commonly employed to model planetary bodies because these objects typically exhibit a nearly spherical shape \citep{Elipe1}. This method becomes less effective when it is applied to objects with irregular geometries, however, because the harmonic expansion might no longer be suitable and convergence issues might arise \citep{1999imda.coll..169R, 2018AdSpR..62.3199J, 2017Ap&SS.362..169L}. 
Traditionally, the gravitational fields of irregular or non-homogeneous mass distributions are modeled using spherical harmonic expansions. However, these series often suffer from high computational costs and poor convergence. These limitations have created a significant bottleneck in current modeling techniques, necessitating the exploration of alternative solutions to the Laplace equation, through the development of closed-form expressions for the potential function \citep{1999ASIC..522..321B}.

A widely used alternative for representing the gravitational field of irregularly shaped bodies is the polyhedron method \citep{1994CeMDA..59..253W, 1996CeMDA..65..313W, WERNER19971071, 2000JGCD...23..466S, tsoulis2001, tsoulis2012}. This model is regarded as highly accurate and does not present convergence challenges \citep{2000JGCD...23..466S}. The polyhedron model has been extensively used in real mission projects to study spacecraft dynamics near asteroids because it is precise. A significant drawback of the polyhedron model is the need for thousands of parameters (vertices and facets) to maintain its high level of accuracy, however. This in turn results in substantial computational demands and extended simulation times. Additionally, the analysis of the general dynamics of spacecraft with respect to the model parameters becomes increasingly complex because the parameters are interrelated and exert a combined effect on the effective gravitational field of irregular bodies.

In light of these limitations, simplified models provide an appealing alternative for studying particle dynamics around complex bodies. They offer a more computationally efficient approach than more elaborate models such as the polyhedron model \citep{1994CeMDA..59..253W} or the mascon model \citep{Geissleretal96, 1996CeMDA..65..313W, Amarante2021}. By reducing computational complexity while maintaining reasonable accuracy, simplified models are valuable tools for investigating the gravitational environment around irregularly shaped celestial bodies.
Simplified models provide an effective approach for approximating the gravitational potential of irregular celestial bodies. These models are particularly useful in celestial mechanics because they help us to overcome significant challenges related to calculating gravitational fields for bodies with nonuniform mass distributions or irregular shapes, such as decreasing time-step numerical integration costs and/or simplifying analytical analyses. Simplified models can often provide a straightforward analytical expression for calculating the gravitational field of irregular bodies \citep{2018RAA....18...84Y}. An additional advantage of using these models is the ease with which the effects of specific parameters on the dynamics around asteroids can be analyzed. This includes, for example, the stability of equilibrium points \citep{2015Ap&SS.356...29Z, 2017Ap&SS.362...61B}, the distribution of stable periodic orbits \citep{2017Ap&SS.362..169L}, and the identification of possible hovering regions \citep{2015RAA....15.1571Y, 2016JGCD...39.1223Z}. Moreover, simplified models are beneficial in orbit design \citep{2017Ap&SS.362..229W, 2017Ap&SS.362...61B} and feedback control strategies \citep{2017Ap&SS.362...27Y}. While specific trajectories can be more accurately analyzed with refined models during the final stages of a mission, a general understanding of the gravitational field through a limited set of parameters proves to be invaluable in the early phases of mission planning.

Several simplified models have been introduced to represent irregular asteroids, as discussed by \cite{1999imda.coll..169R, 2001CeMDA..81..235R, 2003CeMDA..86..131B, Elipe1}, where the dynamics of a particle under the gravitational effect of an asteroid, modeled as a straight segment, were explored. Other studies investigated the motion of a particle around small celestial bodies using simplified models through a simple planar plate \citep{Blesa}, a homogeneous cube \citep{2011Ap&SS.334..357L}, a triaxial ellipsoid \citep{2006SJADS...5..252G}, a rotating homogeneous cube \citep{2011Ap&SS.333..409L}, double and finite straight segments \cite{2014Ap&SS.351...87J}, and a rotating-mass dipole \citep{qui, Kokoriev}. A method for adapting the rotating-mass dipole model to a real asteroid was proposed for the first time by \cite{2015Ap&SS.356...29Z}. This model was subsequently refined by considering the oblateness of the primary bodies \citep{2016Ap&SS.361...14Z, 2016Ap&SS.361...15Z, 2017RAA....17....2Z} and incorporating the dipole-segment model \citep{2018AJ....155...85Z}. From then on, studies for the dipole-segment model were presented by \cite{elipe2021symmetric, 2024AdSpR..74.5687A} with the aim of analyzing the dynamics around irregular asteroids.  In the context of elongated asteroid binary systems, \cite{2017Ap&SS.362...61B} applied the restricted synchronous three-body problem model.

Inspired by \cite{2015Ap&SS.356...29Z}, \cite{2017Ap&SS.362..169L} introduced the rotating-mass tripole model with symmetrical rotation. The authors proposed that small convex bodies are effectively modeled using the mass-tripole model. They demonstrated that by using five parameters (determined with the aid of the polyhedron model), the geometric configuration can be defined, the physical characteristics of a real asteroid can be derived, and its gravitational field can be computed. To extend the work of \cite{2017Ap&SS.362..169L}, \cite{2020RMxAA..56..269D} conducted a semi-analytical investigation into the qualitative dynamics around an asteroid with a convex shape and employed the rotating-tripole model. \cite{2021MNRAS.502.4277S} proposed a simplified model that considered the spatial distribution of the irregular body instead of a distribution in the $xy$-plane to describe the gravitational fields of elongated asteroids. This model provided results with a better accuracy than existing axisymmetric models and the mass-point model at low computational cost. Also for a tripolar system, \cite{boaventura2023analysis} analyzed the dynamics around a simplified tripolar-system model with a segment, based on the study of symmetric periodic orbits, applied to asteroid Holiday. These studies and others demonstrated the value of using simplified models to identify the key parameters that affect the motion of particles in specific asteroid systems.

Numerous models have been proposed to study the gravitational environment around irregular asteroids, with a focus on simplifying the complex dynamics while maintaining a reasonable degree of accuracy. Among these, the dipole-segment model is a commonly used approach for approximating the gravitational field of elongated bodies (e.g., \citep{2018ScChE..61..819Z, Zeng_2018,elipeabadalessandra}).

To address the challenges associated with the gravitational modeling of elongated irregular bodies, we propose the generalized dipole-segment model (GDSM). This model significantly extends the traditional dipole-segment approach by incorporating five key parameters: the force ratio ($k$), the pole masses ($\mu$), the connecting rod mass ($\mu_s$), and the flattening coefficients of the poles ($A_1$ and $A_2$).  
To identify these parameters for our proposed model, we employed nonlinear optimization techniques, which are detailed below. The advantages of the model are outlined below.

By advancing beyond traditional dipole-segment models, which simplify the poles as point masses, the proposed approach introduces distinct pole masses and a massive connecting rod and considers the spheroidal shape of the poles. These refinements represent a meaningful step toward a more realistic representation of the gravitational field of elongated bodies.

This paper is structured as follows. Section \ref{EofM} presents the normalized equation of motion for a particle in the vicinity of elongated asteroids, which are modeled using the Generalized Dipole-Segment model. Section \ref{S2} outlines the method we employed to determine the parameters of the simplified model through a nonlinear optimization approach. In Sect. \ref{S3} the proposed method and model are applied to two irregular asteroids, Arrokoth and Kleopatra, and to comet 103P/Hartley. Section \ref{Section5} compares the Generalized Dipole-Segment model with the Dipole-Segment model by evaluating them against the polyhedron model. This comparison highlights the advantages of the newly proposed model. Section~\ref{section:HO} presents the numerical simulations, in which the equilibrium points computed from the GDSM are used as reference to construct the heteroclinic orbits around asteroids Arrokoth, Kleopatra, and 103P/Hartley. Finally, Sect. \ref{conclusion} summarizes our conclusions.

%--------------------------------------------------------------------
\section{Dynamical equations of the Generalized Dipole Segment}
\label{EofM}

We considered an inertial reference frame denoted as Oxyz. The body was composed of two masses with a spheroid format $m_1$ and $m_2$, connected by a segment with a mass $m_3$ and length $l$, with a constant linear mass density $\rho$. The total mass of the system is $M=m_1+m_2+m_3$. The masses $m_1$ and $m_2$ are assumed to be oblate bodies, meaning that they exhibit a flattened shape. Depending on the value of the oblateness coefficient ($A$), these bodies can either be oblate (flattened at the poles when the coefficient is positive) or prolate (elongated at the poles when the coefficient is negative). The parameter $A_1$ represents the oblateness coefficient of the body with a mass $m_1$, and $A_2$ denotes the oblateness coefficient of the body with a mass $m_2$.

This body rotates uniformly in the xy-plane with a constant angular velocity 
$\omega$ around its center of mass, denoted by O. We refer to this configuration as the Generalized Dipole-Segment (GDS).

The GDS system rotates, and we therefore introduced a synodic reference frame Oxyz such that the axes OZ and Oz were aligned and the body continuously rotated about the Ox-axis. 

We introduced two mass ratios,
\begin{equation}
     \mu=\frac{m_2}{m_1+m_2},~~ \mu_s=\frac{m_3}{m_1+m_2+m_3},
     \label{eq1}
   \end{equation}
which assumed values within the intervals $\mu \in [0, 1]$ and $\mu_s \in [0, 1]$, respectively.

Let (-$l_1$, 0, 0) and ($l_2$, 0, 0), where $l$ = $l_1$+$l_2$ = 1 in canonical units (c.u.) represent the positions of the masses $m_1$ and $m_2$ in the Oxyz reference frame, respectively. $l_1$ and $l_2$ are defined as presented in Eqs. \ref{l1} and \ref{l2},
\begin{equation}
     l_1=\mu(1 - \mu_s) + \frac{\mu_s}{2},
     \label{l1}
   \end{equation}
\begin{equation}
     l_2=(1-\mu)(1 - \mu_s) + \frac{\mu_s}{2}.
     \label{l2}
\end{equation}
When the mass ratio $\mu_s$ is precisely zero, the system simplifies to two basic scenarios: either a single-point mass, or a conventional dipole configuration. Conversely, when $\mu_s$ equals one, the model corresponds to the standard segment configuration.
In the canonical unit system, the masses of bodies $m_1$, $m_2$, and $m_3$ are $m_1$ = (1 - $\mu$)(1 - $\mu_s$), $m_2$ = $\mu$(1 - $\mu_s$), and $m_3$ = $\mu_s$, respectively.

\begin{figure}[h]
    \centering
    \includegraphics[width=0.49\textwidth]{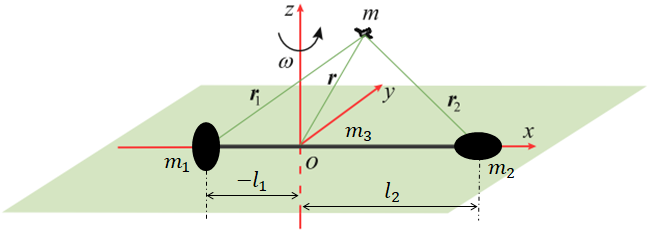}
    \caption{Geometric configuration of the rotating coordinate system for the generalized dipole-segment model, where $m_1$ and $m_2$ represent the masses at the poles, and $m_3$ corresponds to the mass of the central segment. }
    \label{GDSM}
\end{figure}

We considered the Generalized Dipole-Segment model to rotate with a constant angular velocity. The vector equations governing the motion of a test particle affected by the gravitational field of the Generalized Dipole-Segment model are expressed in Eq. \ref{rotating2},
\begin{equation}
\mathbf{\ddot{r}}+2\boldsymbol{\omega}\times\mathbf{\dot{r}}=-\nabla V,
\label{rotating2}
\end{equation}
in the body-fixed frame Oxyz, as depicted in Fig. \ref{GDSM}. The effective potential $V$ is defined in Eq. \ref{V},
\begin{multline}
    V = \omega^2\frac{1}{2} (x^2 + y^2) +  
    \omega^2k \bigg[\frac{(1 - \mu)(1 - \mu_s) }{r_1}\left( 1 + \frac{A_1 \left( r_1^2 - 3z^2 \right)}{2 r_1^4} \right)  \\
    + \frac{\mu (1 - \mu_s) }{r_2} \left( 1 + \frac{A_2 \left( r_2^2 - 3z^2 \right)}{2 r_2^4} \right)
     + \mu_s \log \left( \frac{1 + r_1 + r_2}{-1 + r_1 + r_2} \right)\bigg],
    \label{V}
\end{multline}
where $r_1$ and $r_2$ are given by
\begin{equation}
    r_1=\sqrt{(x+l_1)^2+ y^2+ z^2},
    \label{r1}
\end{equation}
\begin{equation}
    r_2=\sqrt{(x-l_2)^2+ y^2+ z^2}.
    \label{r2}
\end{equation}
$A_i$ ($i$ = 1, 2) represent the oblateness coefficients of the primary bodies, which are defined in Eq. \ref{A},
\begin{equation}
    A=\frac{(\rho_i^e)^2-(\rho_i^p)^2}{5l^2},
    \label{A}
\end{equation}
where the parameter $\rho$ denotes the radius of the primary spheroids. The subscript $e$ corresponds to the equatorial radius of the primary body, and the subscript $p$ refers to its polar radius \citep{2017JPhCS.911a2023S, Abozaid2024}.
The dimensionless scalar parameter $k$ is known as the force ratio, which represents the proportion between gravitational acceleration and centrifugal acceleration, given by Eq. \ref{k} \citep{2021MNRAS.502.4277S},
\begin{equation}
    k=\frac{GM}{\Omega^2l^3},
    \label{k}
\end{equation}
where $\Omega$ and $l$ are the angular velocity and length of the asteroid, respectively, in the international system of units.
The system in Eq. \ref{k} for $k$ = 1 corresponds to the restricted three-body problem. When $k$ $<$ 1, the segment rotates rapidly, whereas $k$ > 1 indicates a slower rotation. As noted by \citet{1994JGCD...17..787P}, cohesive forces are required in cases of rapid rotation to prevent the body from disintegrating as a result of inertial forces that are not counterbalanced by self-gravitation. This suggests that scenarios with $k$ $<<$ 1 are not particularly realistic \citep{2024AdSpR..74.5687A}. 

The gradients of the effective potential in three-dimensional space ($x$, $y$, $z$) are given by Eqs. \ref{x}, \ref{y}, and \ref{z},
\begin{equation}
    \ddot{x}-2\omega\dot{y}=V_x,
    \label{x}
\end{equation}
\begin{equation}
    \ddot{y}+2\omega\dot{x}=V_y,
    \label{y}
\end{equation}
\begin{equation}
    \ddot{z}=V_z,
    \label{z}
\end{equation}
where $V_x$, $V_y$, and $V_z$ denote the partial derivatives of the effective potential $V$ with respect to $x$, $y$, and $z$, respectively.

Equation \ref{ZVC} possesses the well-known Jacobi integral, 
\begin{equation}
    v^2=2V-C,
    \label{ZVC}
\end{equation}
which establishes the permissible region for possible motions. Equation \ref{ZVC} indicates that for a given value of $C$, the velocity $v$ depends on the position of the body within the plane of motion \citep{2017Ap&SS.362...61B}. The integration constant 
$C$ is determined by the initial position and velocity of the particle, as stated by \cite{1963icm..book.....M}. Furthermore, Eq. \ref{ZVC} establishes a connection between the squared velocity and the coordinates of the body with negligible mass in the rotating reference frame, as mentioned by \cite{1970aitc.book.....M}. Consequently, when the integration constant $C$ is numerically obtained from the initial conditions, Eq. \ref{ZVC} allows us to compute the velocity of the body at any point in space. Conversely, for a given velocity, Eq. \ref{ZVC} defines the geometric regions of permissible positions for the body. Notably, when the velocity is set to zero in Eq. \ref{ZVC}, the resulting region corresponds to locations in which the velocity of the particle is also zero, as discussed by \cite{1967torp.book.....S}. Mathematically, this boundary is described by the equation $2V-C=0$, as presented by \cite{1963icm..book.....M}.

\section{Determining the parameters for the generalized dipole-segment model}
\label{S2}

The Generalized Dipole-Segment model was presented in the previous section. This model has unknown parameters that need to be determined to provide a complete set of equations for the system. In the simplified model we considered, the total mass and angular velocity of the asteroids are equivalent to the actual values of these asteroids. The Generalized Dipole-Segment model therefore has five unknown parameters, which are the mass ratio ($\mu$), the mass of the segment ($\mu_s$), the spheroidal shape of the poles ($A_1$ and $A_2$), and the force ratio ($k$).

The unknown parameters were obtained using the locations of the external equilibrium points calculated through the polyhedral approach, following a method similar to that used by \cite{2015Ap&SS.356...29Z, 2016JGCD...39.1223Z, 2018RAA....18...84Y}. The primary objective was to identify the parameters that produce equilibrium points as close as possible to those derived from the polyhedron model. This approach is suitable for asteroids with external equilibrium points. Certain asteroids lack external equilibrium points, however, such as asteroid 1998 KY26. In these cases, an alternative method that minimizes the errors of the effective potential or its gradient can be applied, as suggested by \citep{2018AJ....155...85Z}. The coordinates of the equilibrium points [$\hat{x}_E ,\hat{y}_E,\hat{z}_E$] were calculated as indicated in Eq. \ref{eq31},

\begin{equation}
\label{eq31}
V_x(\hat{x}_E ,\hat{y}_E,\hat{z}_E) = V_y(\hat{x}_E ,\hat{y}_E,\hat{z}_E) = V_z(\hat{x}_E ,\hat{y}_E,\hat{z}_E) = 0.
\end{equation}

The objective was to identify the parameters of the Generalized Dipole-Segment model that minimize the discrepancies between the external equilibrium points of the simplified model and those of the polyhedron model. The procedure we used to determine the parameters for the Generalized Dipole-Segment model is outlined below.

The optimization variables for this model were $X = [\text{$\mu$, $\mu_s$, $A_1$, $A_2$, $k$}]$.
Before we performed the optimization, constraints were set for each variable using the lower bounds [$\text{$\mu$}_{min}$, $\text{$\mu_s$}_{min}$, $\text{$A_1$}_{min}$, $\text{$A_2$}_{min}$, $\text{$k$}_{min}$] and the upper bounds [$\text{$\mu$}_{max}$, $\text{$\mu_s$}_{max}$, $\text{$A_1$}_{max}$, $\text{$A_2$}_{min}$, $\text{$k$}_{max}$] for these parameters. For the nonlinear optimization problem, the performance index is defined as
\begin{multline}
\label{eq32}
\mathbf{J}(\mu, \mu_s, k, A_1, A_2) = \\\sum_{i = 1}^{n}\sqrt{(\hat{x}_{G_i}d^* - x_{P_i})^2 + (\hat{y}_{G_i}d^* - y_{P_i})^2 + (\hat{z}_{G_i}d^* - z_{P_i})^2},
\end{multline}
where  [$\hat{x}_{G_i}, \hat{y}_{G_i}, \hat{z}_{G_i}$] denotes the positions of the equilibrium points obtained from the simplified models (Generalized Dipole-Segment), expressed in canonical units. It is important to note that $l$ can be calculated using Eq. \ref{k}.  [$x_{P_i}, y_{P_i}, z_{P_i}$] represent the coordinates of the equilibrium points (measured in meters) as determined by the polyhedron model. The index $i$ corresponds to the $i$th equilibrium point, and $n$ denotes the total number of external equilibrium points.

The equality constraints ($c_{eq} = 0$) for the Generalized Dipole-Segment model are presented in Eqs. \ref{eq33} and \ref{eq34},
\begin{equation}
\label{eq33}
c_{eq} = \begin{bmatrix}
m_1 + m_2 + m_3 - 1
\end{bmatrix},
\end{equation}
\begin{equation}
\label{eq34}
c_{eq} = \begin{bmatrix}
l_1 + l_2 - 1 
\end{bmatrix}.
\end{equation}

Mathematically, the performance index can be expressed as a constrained minimization, as illustrated in Eq. \ref{eq35},
\begin{equation}
\label{eq35}
min~\mathbf{J}(\mu, \mu_s, k, A_1, A_2)~such~that~ c_{eq} = 0,
\end{equation}
where $\mathbf{J}(\mu, \mu_s, k, A_1, A_2)$ is a function that produces a scalar output, and $c_{eq}$ represents functions that yield vector outputs.
Nonlinear optimization routines implemented in Matlab were used to obtain the optimal solutions that minimize $\mathbf{J}$ \citep{2022AdSpR..70.3362S}. The optimization problem was formulated and solved using a nonlinear programming approach (NLP).

\section{Applications to realistic elongated asteroids}
\label{S3}

In this section, we apply the Generalized Dipole-Segment (GDS) model to two realistic elongated asteroids, KBO 486958 Arrokoth and Kleopatra, and to comet 103P/Hartley (commonly known as Hartley 2). Using an optimization method, we determine the parameters of the GDS for each body.
The main distinction between the dipole-segment (DS) and the Generalized Dipole-Segment (GDS) models lies in the flattening coefficients $A_1$  and $A_2$. When $A_1$ = $A_2$ = 0, the model reduces to the classical dipole-segment formulation. Conversely, when $A_1$ and $A_2$
are nonzero, the Generalized Dipole-Segment model is obtained, which allows us to represent more elongated and asymmetric mass distributions.
We demonstrate the improved performance of the Generalized Dipole-Segment model compared to the dipole-segment model below.

\subsection{Parameters of the elongated asteroids}
\label{S4}

The physical and orbital characteristics of asteroids KBO Arrokoth and Kleopatra and of comet 103P/Hartley were obtained from relevant literature sources and are summarized in Table \ref{table1}. We also employed the polyhedron model to accurately characterize the gravitational fields of these bodies (for further details of this method, we refer to \cite{1994CeMDA..59..253W, 1996CeMDA..65..313W,tsoulis2001,tsoulis2012}).

\onecolumn
\begin{table}
\centering
\caption{Physical and polyhedral parameters of Arrokoth, Kleopatra, and 103P/Hartley.}
\label{table1}
\begin{tabular*}{\textwidth}{c @{\extracolsep{\fill}} *{5}{c}}
\hline
Asteroid & Bulk Density & Rotational Period & Mass  & Vertices \& Facets\\
   & ($g~cm^{-3}$) & (hours) & (kg)  & \\
\toprule
\hline
Arrokoth & 0.235 & 15.938 & $1.58\times10^{15}$ & 53,755 \& 107,506 \\
Kleopatra & 4.9 & 5.385 & $4.68\times10^{18}$ & 1,148 \& 2,292 \\
103P/Hartley & 0.3 & 18.0 & $2.43\times10^{11}$ & 16,022 \& 32,040 \\
\hline
\end{tabular*}
\bigskip
\bigskip

\caption{Positions of equilibrium points for Arrokoth, Kleopatra, and 103P/Hartley.}
\label{table2}
\resizebox{1.\textwidth}{!}{
\begin{tabular}{|p{2.8cm}|p{2.8cm}|p{3.8cm}|p{3.8cm}|p{3.8cm}|p{3.8cm}|}
\toprule
\hline
Asteroid & Model & $E_1$ (km) & $E_2$ (km) & $E_3$ (km) & $E_4$ (km)\\
\hline
\multirow{3}{*}{Arrokoth} 
 & Polyhedron & (19.52, 0.180, -0.061) & (-2.677, 14.13, 0.036) & (-21.29, -0.068, -0.113) & (-2.756, -14.05, -0.001) \\
 & GDSM        & (19.5332, 0, 0) & (-2.75476, 14.0577, 0) & (-21.3064, 0, 0) & (-2.75476, -14.057, 0) \\
 & DSM         & (19.51, 0, 0) & (-1.5152, 13.8118, 0) & (-21.3029, 0, 0) & (-1.5152, -13.8118, 0) \\
\hline
\multirow{3}{*}{Kleopatra} 
 & Polyhedron & (175.3, 0.593, 0.617) & (-0.211, 129.0, 0.238) & (-175.4, 0.898, -0.686) & (-0.034, -129.0, 0.184) \\
 & GDSM        & (175.341, 0, 0) & (-0.1116, 129.041, 0) & (-175.419, 0, 0) & (-0.1116, -129.041,0) \\
 & DSM         & (175.313, 0,0) & (-0.10585,129.042,0) & (-175.455,0,0) & (-0.105857, -129.042, 0) \\
\hline
\multirow{3}{*}{103P/Hartley} 
 & Polyhedron & (1.511, -0.005, -0.004) & (0.141, 1.074, -0.003) & (-1.409, -0.037, 0.011) & (0.137, -1.073, -0.003) \\
 & GDSM        & (1.5113,0,0) & (0.1389, 1.0741,0) & (-1.4096,0,0) & (0.1389, -1.0741,0) \\
 & DSM         & (1.51094,0,0) & (0.1381,1.0721,0) & (-1.3781,0,0) & (0.1381,-1.0721,0) \\
\hline
\bottomrule
\end{tabular}}
\bigskip
\end{table}

\begin{table}
\centering
\caption{Optimization results for the GDSM.}
\label{table4}
\bigskip
\begin{tabular}{|l||c|c|c|c|c|c|c|c|}% {\textwidth}{c @{\extracolsep{\fill}}*{5}{c}}
Asteroid & $\mu$ & $\mu_s$ & $k$ & $A_1$ & $A_2$ & \textbf{J} & \textbf{$J_{max}$} & \textbf{$J_{min}$}\\
 & [c. u.] & [c. u.] &  & [c. u.] & [c. u.] & [km] & \% & \% \\
\hline
Arrokoth & 0.9587 & 0.5712 & 0.2754 & -0.022627 & 0.0300 & 0.4414 & 0.7705 & 0.0148\\
Kleopatra & 0.5008 & 0.4603 & 1.0420 & 0.0444 & 0.0445 & 2.4495 & 0.4741 & 0.0844\\
103P/Hartley & 0.3513 & 0.1944 & 0.8747 & 0.0379 & 0.0364 & 0.0539 & 3.0792 & 0.2970\\
\end{tabular}

\bigskip
 \caption{Optimization results for the DSM.}
 \label{table5}
\begin{tabular}{|l||c|c|c|c|c|c|c|c|}% {\textwidth}{c @{\extracolsep{\fill}}*{5}{c}}
Asteroid & $\mu$ & $\mu_s$ & $k$ & $A_1$ & $A_2$ & \textbf{J} & \textbf{$J_{max}$} & \textbf{$J_{min}$}\\
 & [c. u.] & [c. u.] &  & [c. u.] & [c. u.] & [km] & \% & \%\\
\hline
Arrokoth & 0.7324 & 0.5230 & 0.3491 & 0 & 0 & 2.1811 & 7.9551 & 0.0296\\
Kleopatra & 0.5065 & 0.7723 & 0.3694 & 0 & 0 & 2.4507 & 0.4274 & 0.0993\\
103P/Hartley & 0.1551 & 0.5424 & 0.5000 & 0 & 0 & 0.0662 & 3.2995 & 0.3336\\
\end{tabular}
\end{table}

\twocolumn

\subsection{Equilibrium points}
We employed the polyhedron model as described by \cite{1994CeMDA..59..253W, 1996CeMDA..65..313W,tsoulis2001,tsoulis2012}. Using this approach, we determined the positions of the equilibrium points numerically by solving the condition
\begin{equation}
\nabla V = \mathbf{0}.
\label{equilibrium-points}
\end{equation}
Using the gravitational potential from polyhedron model (Eq. \eqref{eq:potencial_split} together with the \textsc{Minor-Equilibria} package\footnote{\url{https://github.com/a-amarante/minor-equilibria-nr}} \citep{amarante2020,minor-equilibria}, we computed the equilibrium points by solving the force balance defined by Eq. \eqref{equilibrium-points}.

The results are presented in Table \ref{table2}. We emphasize that only the external equilibrium points were considered because the internal points have no physical significance.

To perform the simulations, it was essential to establish the boundary constraints for each asteroid we studied using the generalized dipole-segment model. These constraints are defined below.

\begin{enumerate}
\item Generalized dipole-segment model

For each asteroid, [$\mu_{min}$, $\mu_{max}$] were set to [0.001, 0.999] kg, [$\mu_{s,~min}$, $\mu_{s,~max}$] were set to [0.001, 0.999], [$k_{min}$, $k_{max}$] were set to [0.1, 9], [$k_{min}$, $k_{max}$] were set to [0, 9], [$A_{1,min}$, $A_{1,max}$] were set to [-4, 4], and [$A_{2,~min}$, $A_{2,~max}$] were set to [-4, 4].

\item Dipole-segment model

For each asteroid, [$\mu_{min}$, $\mu_{max}$] were set to [0.001, 0.999] kg, [$\mu_{s,~min}$, $\mu_{s,~max}$] were set to [0.001, 0.999], [$k_{min}$, $k_{max}$] were set to [0.1, 9], [$k_{min}$, $k_{max}$] were set to [0, 9], [$A_{1,min}$, $A_{1,max}$] were set to [0, 0], and [$A_{2,~min}$, $A_{2,~max}$] were set to [0, 0].
\end{enumerate}

The boundary constraints for the geometric parameters $\mu$ and $\mu_s$ were determined based on the classical canonical unit system definition, where the mass ratios cannot exceed unity because their sum must satisfy the condition $\mu + \mu_s = 1$.

The parameter $k$ was initially defined within the range $[0.1, 9]$. When the optimal solution obtained from the simulations approached the upper limit of this range, the interval was adjusted to $[5, 15]$, and the optimization was performed again. This process was repeated iteratively until the optimal value of $k$ was well centered within the defined range.  

A similar approach was applied to the flattening coefficients $A_1$ and $A_2$, which were initially constrained within the interval $[-4, 4]$. When the optimization results indicated values close to the boundaries, the range was adjusted accordingly. This adaptive procedure ensured that the final parameter values remained well within the specified limits.  

The initial guesses for the GDSM and DSM models were chosen as described below.

\begin{enumerate}
\item GDSM

For Arrokoth, we used [$\mu$, $\mu_s$, $k$, $A_1$, $A_2$] = [0.1, 0.8, 5, 0.1, 0.1], for Kleopatra, we used [$\mu$, $\mu_s$, $k$, $A_1$, $A_2$] = [0.3, 0.3, 5, 0.1, 0.1], and for 103P/Hartley, we used [$\mu$, $\mu_s$, $k$, $A_1$, $A_2$] = [0.6, 0.54, 0.5, 0.1, 0.1].

\item DSM

For Arrokoth, we used [$\mu$, $\mu_s$, $k$, $A_1$, $A_2$] = [0.8, 0.54, 8, 0, 0], for Kleopatra, we used [$\mu$, $\mu_s$, $k$, $A_1$, $A_2$] = [0.3, 0.3, 5, 0, 0], and for 103P/Hartley, we used [$\mu$, $\mu_s$, $k$, $A_1$, $A_2$] = [0.6, 0.54, 0.5, 0, 0].
\end{enumerate}

The optimal values of the parameters $\mu$, $\mu_s$, $k$, $A_1$, and $A_2$, determined using the optimization method described above, are presented in Tables \ref{table4} and \ref{table5}. These tables also display the performance indices $J_{\text{max}}$ and $J_{\text{min}}$. The performance indices were computed after we determined the optimal parameters, denoted $J$. Specifically, the indices $J_{\text{max}}$ and $J_{\text{min}}$ represent the maximum and minimum relative errors for a single equilibrium point and were defined as follows.

$J_{\text{max}}$ refers to the maximum relative error between the equilibrium points calculated using the simplified model and the actual equilibrium points located within the high-fidelity model (the polyhedron model). For four equilibrium points, the error in the location of each equilibrium point was computed for the four external equilibrium points. The largest of these relative errors was designated $J_{\text{max}}$, and the smallest was identified as $J_{\text{min}}$. 
The parameter $J_{\text{max}}$ and $J_{\text{min}}$ depends on the optimization variables $(\mu, \mu_s, k, A_1, A_2)$, and it was defined as shown in Eqs. \ref{eq36} and \ref{eq37},
\begin{multline}
\label{eq36}
\mathbf{J_{max}} =  max\left(\frac{\sqrt{(\hat{x}_{G_i}d^* - x_{P_i})^2 + (\hat{y}_{G_i}d^* - y_{P_i})^2 + (\hat{z}_{G_i}d^* - z_{P_i})^2}}{L}\right) \\ \times 100\%,~~~~  i = 1, 2, 3, 4,
\end{multline}

\begin{multline}
\label{eq37}
\mathbf{J_{min}} = min\left(\frac{\sqrt{(\hat{x}_{G_i}d^* - x_{P_i})^2 + (\hat{y}_{G_i}d^* - y_{P_i})^2 + (\hat{z}_{G_i}d^* - z_{P_i})^2}}{L}\right) \\ \times 100\%, ~~~~ i = 1, 2, 3, 4.
\end{multline}

The results show that the values of $J$ are consistently lower when the GDSM is used, which indicates a more accurate representation of the system. For highly elongated bodies with relatively uniform extremities, that is, without significant variations in shape at their poles, the minimized $J$ values yielded comparable results for both models, however. This is particularly evident for asteroid Kleopatra, where the mass parameter $\mu_s$ is comparable to the mass at the extremities.  

On the other hand, when the body had a noticeable bulge at one of its extremities and deviated from a nearly uniform elongated shape, the GDSM proved to be significantly more effective than the DSM. In these cases, the values of the flattening coefficients $A_1$ and $A_2$ are substantial, which reinforces the insight that the Generalized Dynamical Shape Model captures the asymmetry of the system better.  

The accuracy of the GDSM becomes even more evident for the mass distribution within the system. When the body presents a pronounced protrusion at its poles, most of the mass is concentrated at these extremities, which are typically asymmetric and not perfectly spherical. This structural characteristic enhances the performance of the GDSM because it captures the gravitational effects arising from these irregularities better.  

Conversely, for bodies with a more uniform shape, where the poles do not exhibit significant protrusions, the mass distribution along the central structure and at the extremities tends to be on the same order of magnitude. In other words, the mass of the elongated central region is comparable to the mass that is concentrated at the poles. As a result, the system behaves in a nearly homogeneous manner, which leads to comparable optimized values of $J$ for both models, with $J$ remaining below unity. The DSM then also provides satisfactory results, although the GDSM still proves to be the more precise approach.    

A clear example is the case of KBO Arrokoth and comet 103P/Hartley, where the mass parameter $\mu_s$ is small, which leads to a mass distribution that is concentrated at the poles. Because these poles are not spherical, the GDSM significantly outperforms the DSM in accurately modeling the gravitational field of the system.

\section{Comparison between the simplified and the polyhedron models}
\label{Section5}

% \subsection{Polyhedron model}
The polyhedron approach represents an irregularly shaped body with a uniform density ($\rho$) as a collection of tetrahedra bounded by triangular facets. Based on this geometric description, the gravitational potential and the gradient of gravitational potential can be expressed as (\citealt{tsoulis2001}: \citealt{tsoulis2001})
\begin{equation}
\begin{split}
U(x_1, x_2, x_3) = -\frac{G \rho}{2} \sum_{p=1}^{n} \sigma_p h_p \biggl[\sum_{q=1}^{m} \sigma_{pq} h_{pq} L_{N_{pq}} +\\ h_p \sum_{q=1}^{m} \sigma_{pq} A_{N_{pq}} + \sin(gA_p) \biggr],
\end{split}
\label{eq:potencial_split}
\end{equation}
\begin{equation}
\begin{split}
- \frac{\partial U(x_1, x_2, x_3)}{\partial x_i} = -G \rho \sum_{p=1}^{n} \cos(\mathbf{N}_p, \mathbf{e}_i) \biggl[ \sum_{q=1}^{m} \sigma_{pq} h_{pq} L_{N_{pq}} +\\ h_p \sum_{q=1}^{m} \sigma_{pq} A_{N_{pq}} + \sin(gA_p) \biggr], \\\quad (i = 1, 2, 3).
\end{split}
\label{eq:atracao_split}
\end{equation}
Auxiliary functions are given by
\begin{equation}
L_{N_{pq}} = \ln\left( \frac{s_{2pq} + l_{2pq}}{s_{1pq} + l_{1pq}} \right),
\label{eq:lnpq}
\end{equation}
\begin{equation}
A_{N_{pq}} = \arctan\left( \frac{h_p s_{2pq}}{h_{pq} l_{2pq}} \right) - \arctan\left( \frac{h_p s_{1pq}}{h_{pq} l_{1pq}} \right),
\label{eq:anpq}
\end{equation}
\noindent where the parameters of Eqs. \eqref{eq:potencial_split}-\eqref{eq:anpq} were defined by \citet{tsoulis2001} in Sec \ref{Section5}.

The results indicate that the equilibrium points computed using the GDSM agree better with those derived from the polyhedron model than the points obtained with the DSM, as shown in Table \ref{table2}.
This constitutes the first evidence that supports that the GDSM approximates the reference model better. Next, we analyzed the classifications of the equilibrium points for the three asteroids we studied. Our analysis is consistent with the analysis conducted using the polyhedron model. We then compared the relative errors between the gradients of the pseudo-potential computed with the GDSM and those obtained from the polyhedron model in order to assess the validity of the simplified formulation. Because the gradient of the pseudo-potential is a vector quantity, with components associated with the partial derivatives in $x$, $y$, and $z$, the comparison was performed using the magnitude of this vector at each sampled spatial location. The relative error was then calculated as shown in Eq. \ref{realtive},
\begin{equation}
\varepsilon_i = \left( \frac{\nabla V^{\text{GDSM}}_i - \nabla V^{\text{poly}}_i}{\nabla V^{\text{poly}}_i} \right) \times 100\%,
\label{realtive}
\end{equation}
where \( \nabla V^{\text{GDSM}}_i \) and \( \nabla V^{\text{poly}}_i \) denote the gradients of the pseudo-potential evaluated at the same spatial position in the rotating reference frame using the GDSM and the polyhedron model, respectively.
The mean and maximum values of \( \varepsilon_i \) over all sampled points were then analyzed 
to assess the accuracy of the simplified model.
The gradient of the gravitational potential from the polyhedron model was computed with the package \textsc{Minor-Gravity} \footnote{\url{https://github.com/a-amarante/minor-gravity}} \citep{amarante2020,minor-gravity} using Eqs. \eqref{eq:atracao_split}-\eqref{eq:anpq}.

\begin{figure}
\centering\includegraphics[width=1\linewidth]{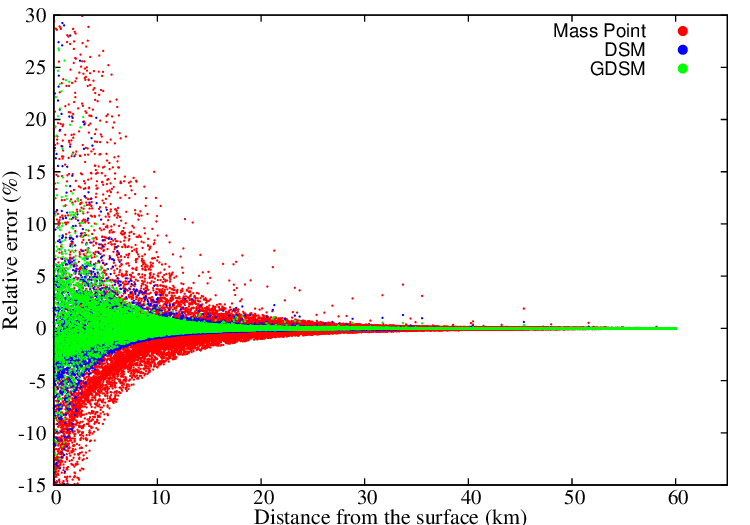}
\caption{Relative error of the gradient of the pseudo-potential for KBO Arrokoth.}
\label{figure3}
\end{figure}

\begin{figure}
\centering\includegraphics[width=1\linewidth]{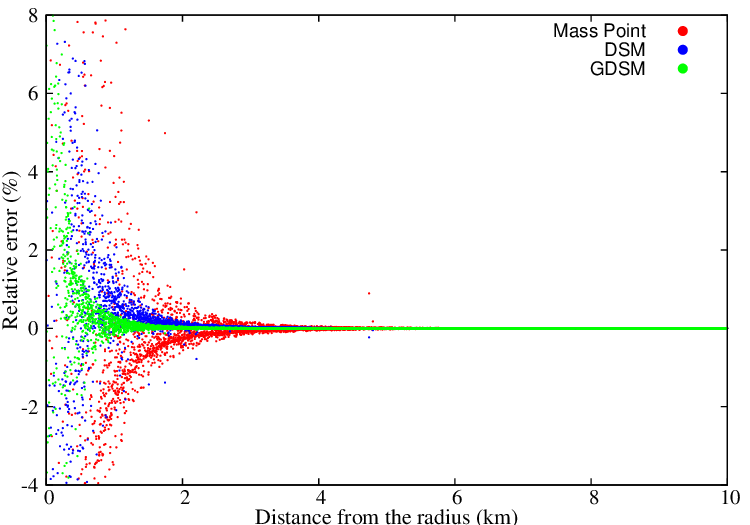}
\caption{Relative error of the gradient of the pseudo-potential for comet 103P/Hartley.}
\label{figure4}
\end{figure}

\begin{figure}
\centering\includegraphics[width=1\linewidth]{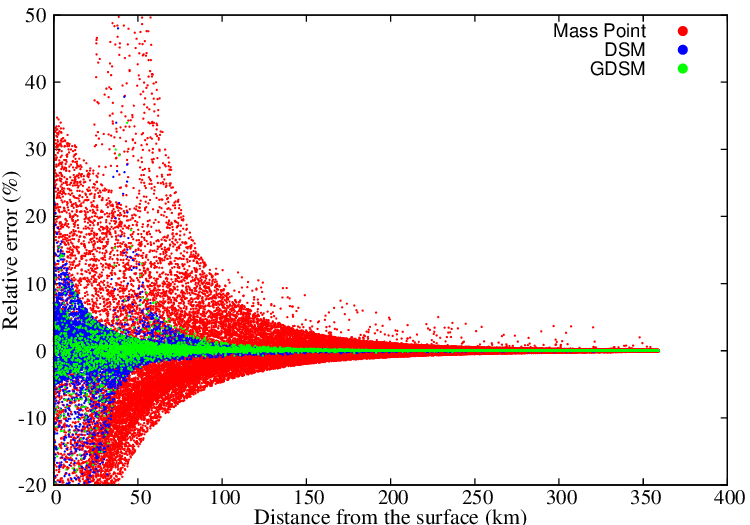}
\caption{Relative error of the gradient of the pseudo-potential for asteroid Kleopatra.}
\label{figure5}
\end{figure}

Figures \ref{figure3}, \ref{figure4}, and \ref{figure5} show the relative errors of the gradients of the pseudo-potential between the simplified models and the polyhedron model for asteroid KBO Arrokoth, asteroid Kleopatra, and comet 103P/Hartley, respectively. The plots display the relative error obtained when modeling the body as a point mass (red), with the DSM (blue), and with the GDSM (green).

As illustrated in Fig. \ref{figure3}, when the spacecraft approaches KBO Arrokoth, particularly at distances shorter than 20 km from its center, the point-mass approximation becomes inadequate because it produces significant relative errors in the computed gradient of the pseudo-potential. When the body is represented by either the DSM or the GDSM formulation, however, the resulting gradient of the pseudo-potential is strong consistent with the reference model. In regions far from the surface of the asteroid, the relative error obtained using the point-mass model, the DSM and the GDSM becomes comparable, indicating that the irregular shape of the body has little effect at large distances. In regions close to the surface of KBO Arrokoth, the relative error in the gradient of the pseudo-potential computed using the GDSM remains at about 7\%. In contrast, the DSM presents errors of almost 15\% under the same conditions. This significant difference highlights the superiority of the GDSM in accurately representing the dynamical environment near the asteroid. Consequently, for spacecraft dynamics analyses in the vicinity of KBO Arrokoth, the GDSM provides a more realistic and reliable approximation.

Figure \ref{figure4} illustrates the relative error in the magnitude of the gradient of the pseudo-potential for three different gravitational representations of comet Harltey2 when compared to the polyhedron model. The point-mass approximation (red) differs most, particularly near the surface, where its simplified structure fails to capture the highly irregular mass distribution of the body. The dipole-segment model (DSM), shown in blue, reduces these errors, but still deviates noticeably between 0 and 2 km from the comet surface. In contrast, the errors of the generalized dipole-segment model (GDSM), shown in green, remain nearly zero at distances greater than 0.5 km from the surface. The GDSM curve remains tightly concentrated around the reference line and agrees excellently with the polyhedron model. This behavior highlights the improved capacity of the GDSM to reproduce the true gravitational environment of elongated bodies while retaining the low computational cost of simplified models.

For the case of asteroid Kleopatra (Fig. \ref{figure5}),  a similar trend is observed. At large distances from the surface, all three formulations, the mass-point, the DSM, and the GDSM, provide comparable results, with relative errors that rapidly diminish with increasing distance. As the evaluation points approach the surface of the elongated body, however, the differences between the models become more pronounced. In this regime, the GDSM formulation is a clear improvement over the mass-point and the DSM representations because it maintains significantly lower errors and is markedly superior in capturing the local gravitational environment near the surface.

The results demonstrate that the DSM already performs well near the surfaces of irregular bodies overall. It is a clear improvement over the point-mass model, whose errors become large under strong geometric asymmetries. The GDSM consistently achieves a noticeably higher level of accuracy, however, particularly in the regions in which the gravitational field varies sharply due to complex shape features. The GDSM preserves low errors across a wider range of distances and represents surface-driven irregularities with significantly better fidelity.

\section{Heteroclinic orbits}
\label{section:HO}

Heteroclinic orbits are trajectories in phase space that connect two different equilibrium points of a dynamical system. In the case of the Restricted Three-Body Problem (RTBP; \citealt{gomez1988homoclinic}), the dipole-segment model \citep{elipe2021symmetric, 2024AdSpR..74.5687A}, and similarly for the GDSM, heteroclinic orbits connect the triangular equilibrium points $E_2$ and $E_4$ when these points are unstable. 

The heteroclinic orbits represent a natural path allowed by the system dynamics in conservative problems to change from one configuration to the next without the need for continuous propulsion. They are important for planning low-fuel missions because they exploit only the natural dynamics of the system.

In the GDSM, the dynamical system is described by the equations of motion, Eqs. \ref{x}, \ref{y}, and \ref{z}, and the equilibrium points are determined according to Eq.\eqref{eq31} and Table ~\ref{table2}. We therefore observe that the triangular points are unstable, as indicated in Table ~\ref{tab:stabpoints}.

\begin{table}[H]
\caption{Stability of the triangular equilibrium points for Arrokoth, 103P/Hartley, and Kleopatra.}
\centering
\renewcommand{\arraystretch}{1.3}
\setlength{\tabcolsep}{6pt}
\begin{tabular}{c|c|c|c}
\hline
Body & Point & Eigenvalues & Stability \\ \hline
Arrokoth      & $E_2$ & $\pm 0.659974 \;\pm 0.932612 i$ & unstable \\ 
              & $E_4$ & $\pm 0.659974 \;\pm 0.932612 i$ & unstable \\ \hline
Kleopatra     & $E_2$ & $\pm 0.652388 \;\pm 0.933019 i$ & unstable \\ 
              & $E_4$ & $\pm 0.652388 \;\pm 0.933019 i$ & unstable \\ \hline
103P/Hartley  & $E_2$ & $\pm 0.619332 \;\pm 0.917642 i$ & unstable \\ 
              & $E_4$ & $\pm 0.619332 \;\pm 0.917642 i$ & unstable \\ \hline
\end{tabular}

\label{tab:stabpoints}
\end{table}

If $p$ is an equilibrium point of a vector field $X$ defined on a certain manifold $M$, knowing that the stable manifolds ($W^s$) are the set of states that eventually approach the point and the unstable manifolds ($W^u$) are the set of states that came from the point before, then we have
\begin{equation}
W^{s}(p) = \Big\{\, \mathbf{x} \in M \;\Big|\;
\lim_{t \to +\infty} \phi_t(\mathbf{t,x}) = p \,\Big\}
\label{eq:ws}
\end{equation}
\begin{equation}
W^{u}(p) = \Big\{\, \mathbf{x} \in M \;\Big|\;
\lim_{t \to -\infty} \phi_t(\mathbf{t,x}) = p \,\Big\},
\label{eq:wu}
\end{equation}
where $\phi_t$ is the flux of the dynamical system.
Now, we have the mathematical conditions for the heteroclinic orbit, in the GDSM, for $p=E_2$ and $p=E_4$,\\

\centerline{{$W^u(E_2) \cap W^s(E_4)$ or $W^s(E_2) \cap W^u(E_4)$},}

\noindent that is, a heteroclinic orbit is characterized by the intersection of the two manifolds, and any intersection point corresponds to an initial condition that generates a trajectory.

According to \cite{gomez1988homoclinic}, the intersections between the manifolds on the $x$-axis, which provide the initial condition of the orbit, can be obtained in the ($x,\dot x$) plane because $y=0$ and $\dot y(x,C)$, with $C$ being the Jacobi constant at the triangular equilibrium point. The graphs define the manifolds at the first crossing with the $x$-axis.

Next, we present the ($x,\dot x$) plane for each body we studied for stable and unstable manifolds (see the vertical dashed red lines that limit the size of the body), following by the respective heteroclinic orbits. Table \ref{tab:cis} presents the initial conditions $(x,C)$ of each orbit for each body.

\begin{table}[H]
\caption{Initial conditions $(x,C)$ of the heteroclinic orbits of Arrokot, Kleopatra, and 103P/Hartley.}
\centering
\renewcommand{\arraystretch}{1.3}
\setlength{\tabcolsep}{6pt}
\begin{tabular}{c|c|c|c}
\hline
       & Arrokoth     & Kleopatra  & 103P/Hartley \\ \hline
 $U_n$ &  $(-0.81449,2.0622)$  & $(-0.75641,2.8831)$ &  $(-0.62442, 2.5735)$ \\
 $U_p$ &  $( 1.68709,2.0622)$  & $( 1.9274, 2.8831)$ &  $(1.79235, 2.5735)$  \\ 
 $S_n$ &  $(-1.61398,2.0622)$  & $(-1.92658,2.8831)$ &  $(-1.8392,2.5735)$ \\ 
 $S_p$ &  $( 0.61573,2.0622)$  & $(0.75496, 2.8831)$ &  $(0.82165, 2.5735)$  \\ \hline

\end{tabular}

\label{tab:cis}
\end{table}

\begin{figure}
\centering\includegraphics[width=1\linewidth]{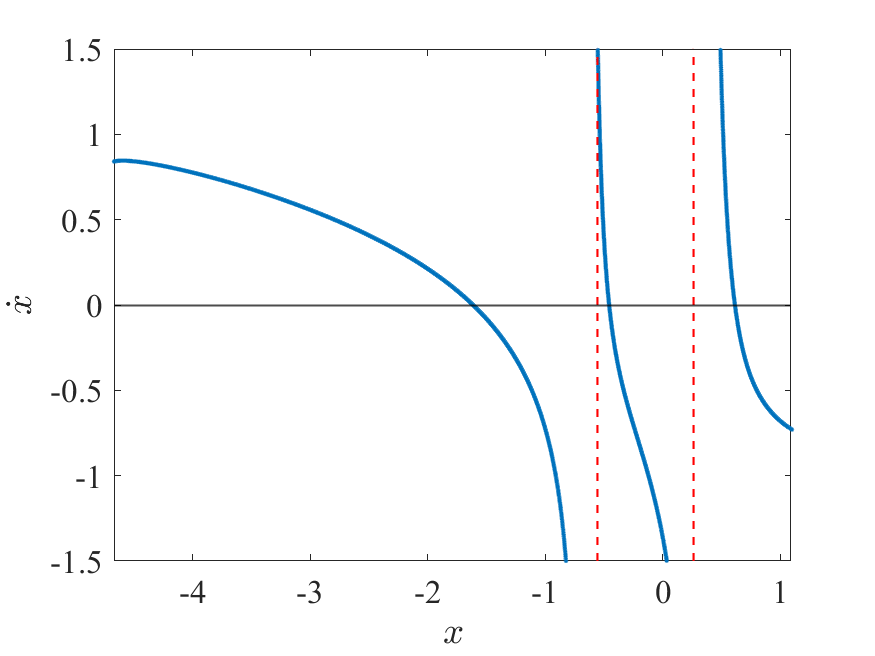}
\caption{Stable manifold ($W^s(E_2)$) at the first crossing for Arrokoth.}
\label{WsArrokoth}
\end{figure}

\begin{figure}
\centering\includegraphics[width=1\linewidth]{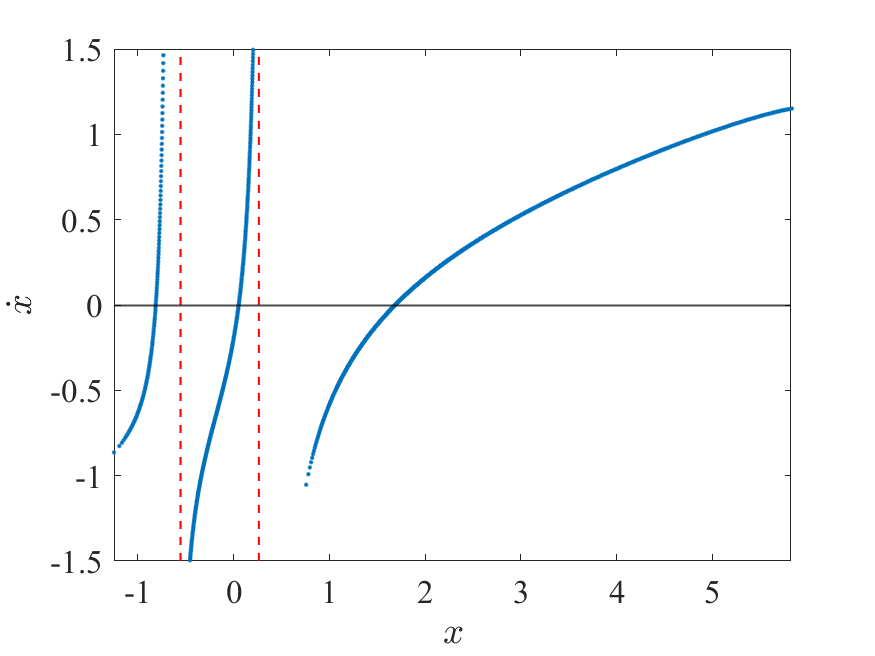}
\caption{Unstable manifold ($W^u(E_2)$) at the first crossing for Arrokoth.}
\label{WuArrokoth}
\end{figure}

Fig. \ref{WsArrokoth} shows for the stable manifold and Fig. \ref{WuArrokoth} for the unstable manifold that they have two initial conditions of heteroclinic orbits for Arrokoth at $\dot x=0$ because the region between the vertical dashed red lines lies within the body. The conditions allocated there are physically impossible.

\begin{figure}
\centering\includegraphics[width=1\linewidth]{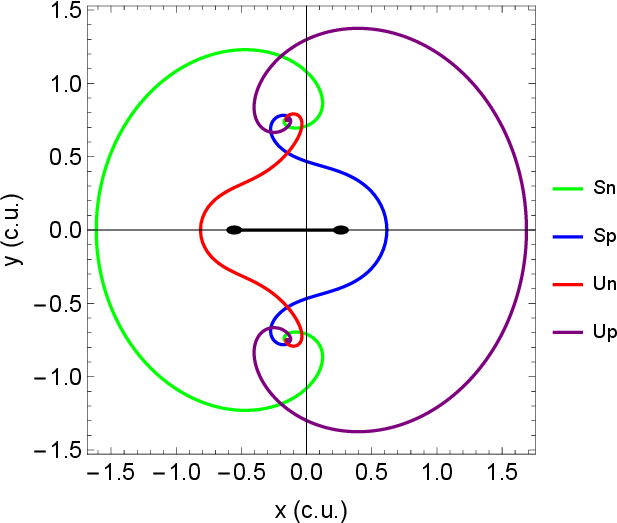}
\caption{Four heteroclinic orbits of Arrokoth, $U_n, U_p, S_n,and  S_p$.}
\label{HOArrokoth}
\end{figure}

Four heteroclinic orbits were found: $U_n$ and $U_p$ are the unstable manifolds that leave $E_2$ and arrive at $E_4$, and $S_n$ and $S_p$ are their respective stable manifolds, which leave $E_4$ and arrive at $E_2$, as shown in Fig. \ref{HOArrokoth}.

\begin{figure}
\centering\includegraphics[width=1\linewidth]{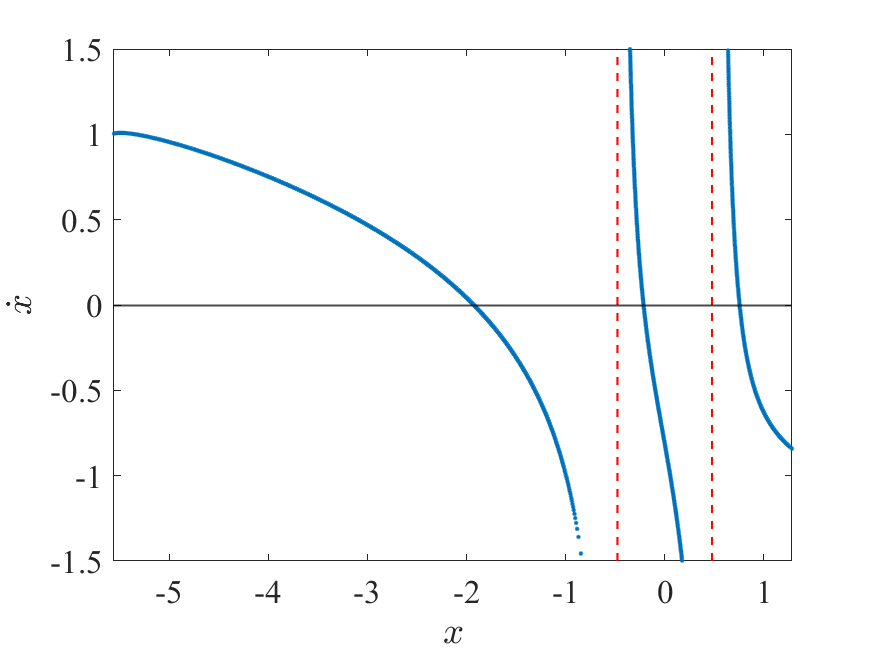}
\caption{Stable manifold ($W^s(E_2)$) at the first crossing for Kleopatra.}
\label{WsKleopatra}
\end{figure}

\begin{figure}
\centering\includegraphics[width=1\linewidth]{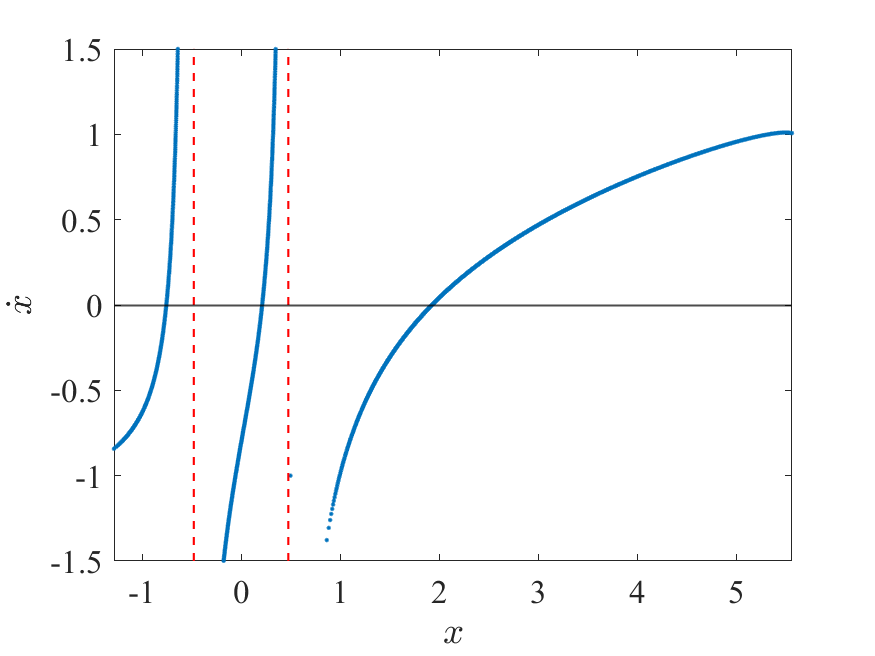}
\caption{Unstable manifold ($W^u(E_2)$) at the first crossing for Kleopatra.}
\label{WuKleopatra}
\end{figure}

In the sequence, we searched the results for asteroid Kleopatra. The stable and unstable manifolds are presented in Figs. \ref{WsKleopatra} and \ref{WuKleopatra}.

\begin{figure}
\centering\includegraphics[width=1\linewidth]{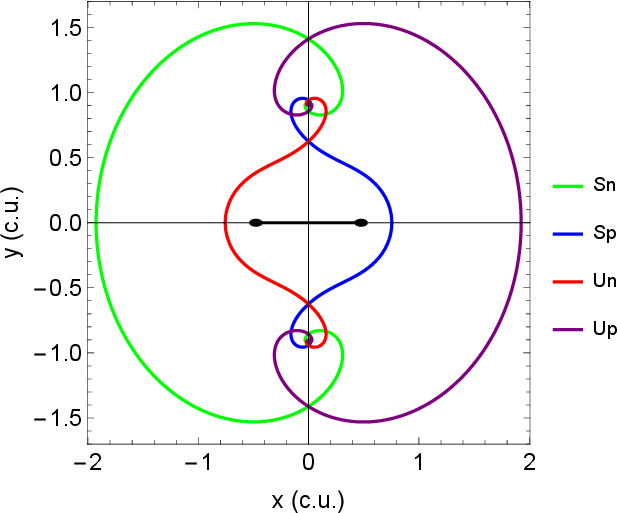}
\caption{Four heteroclinic orbits of Kleopatra: $U_n, U_p, S_n, and S_p$.}
\label{HOKleopatra}
\end{figure}

Figure \ref{HOKleopatra} shows the four heteroclinic orbits that were found. The result is consistent with \cite{2024AdSpR..74.5687A} for the dipole-segment model (DSM). The variation in the shape of the extreme poles of the body from circular (DSM) to ellipsoidal (GDSM) caused small variations in the initial conditions of the orbits \ref{tab:cis} and their magnitudes.

According to \cite{2024AdSpR..74.5687A}, the Kleopatra heteroclinic orbits combined in stable-unstable pairs form symmetric periodic orbits of the families (see \cite{2024AdSpR..74.5687A} for the shape of the families),
\begin{align} \nonumber
\text{$S_n-U_p$} &\;\;\rightarrow\;\; \text{Family $l$} \\ \nonumber
\text{$S_n-U_n$} &\;\;\rightarrow\;\; \text{Family $b$} \\ \nonumber
\text{$S_p-U_p$} &\;\;\rightarrow\;\; \text{Family $a$} \\ \nonumber
\text{$S_p-U_n$} &\;\;\rightarrow\;\; \text{Family $k$} \nonumber.
\end{align}

\begin{figure}
\centering\includegraphics[width=1\linewidth]{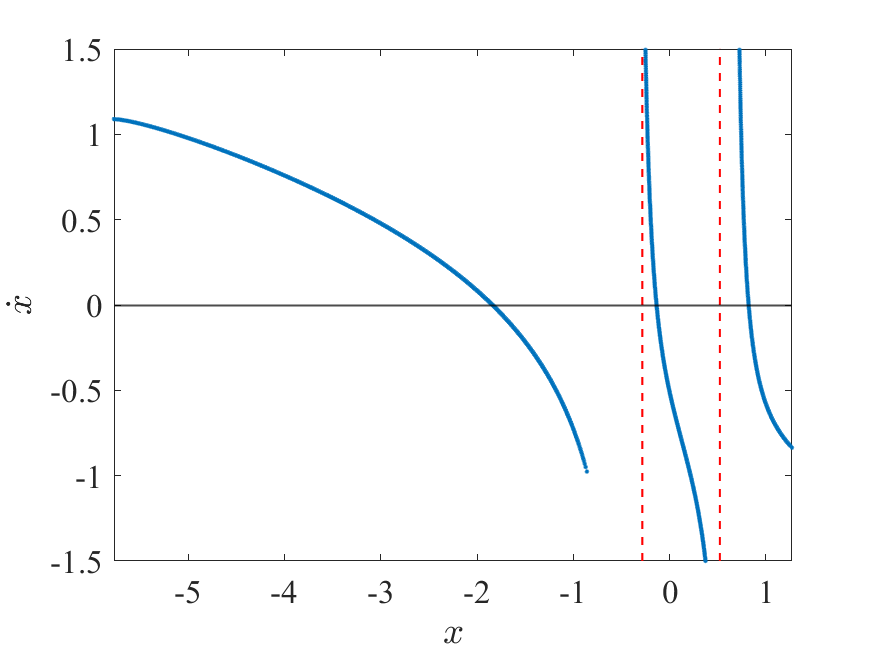}
\caption{Stable manifold ($W^s(E_2)$) at the first crossing for 103P/Hartley.}
\label{WsHartley}
\end{figure}

\begin{figure}
\centering\includegraphics[width=1\linewidth]{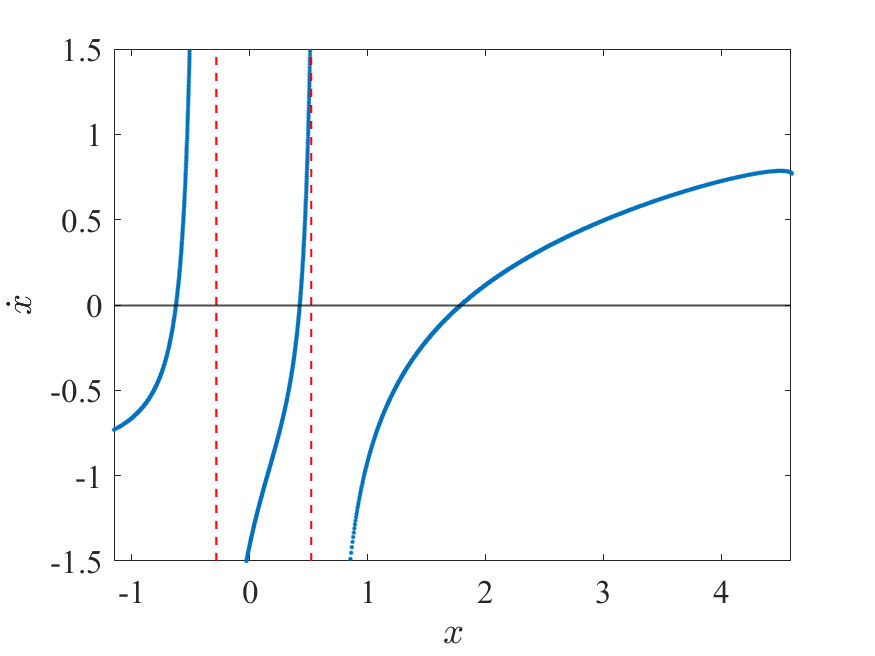}
\caption{Unstable manifold ($W^u(E_2)$) at the first crossing for 103P/Hartley.}
\label{WuHartley}
\end{figure}

\begin{figure}
\centering\includegraphics[width=1\linewidth]{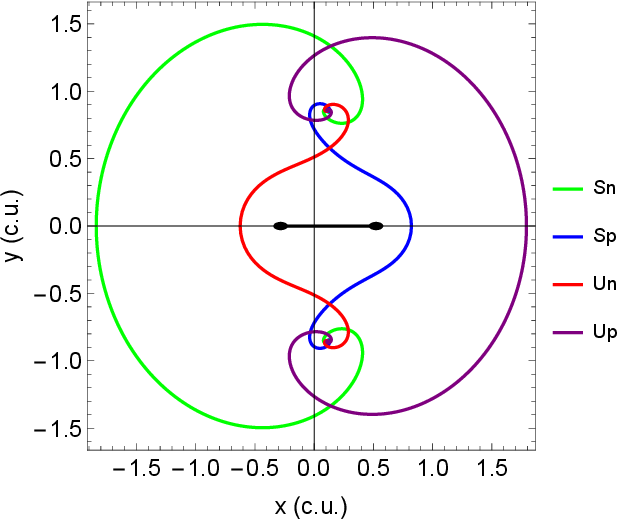}
\caption{Four heteroclinic orbits of 103P/Hartley: $U_n, U_p, S_n, and S_p$.}
\label{HOHartley}
\end{figure}

Finally, we have the stable and unstable manifolds at the first crossing with the $x$-axis for 103P/Hartley (see Figs. \ref{WsHartley} and \ref{WuHartley}) and their respective heteroclinic orbits (Fig. \ref{HOHartley}).

The heteroclinic orbits for the three bodies modeled by GDSM are similar, with differences in magnitudes and initial conditions. Orbits with smaller amplitudes are found for bodies whose distance from the origin to the extremity of the body is greater, and smaller amplitudes for the opposite (Arrokoth and 103P/Hartley). For Kleopatra, whose distance is more balanced and ranges from (-0.4778, 0.4762), the orbits tend to have a balanced amplitude as well.

\section{Conclusions}
\label{conclusion}

We developed the GDSM to represent the gravitational field of elongated and irregular bodies. The main improvement in the GDSM lies in the representation of the flattened shape of the poles, which differs from the DSM and allows us to approximate the mass distribution of the body better. This improvement leads to a greater reliability, particularly in regions close to the surface. We applied the model to the KBOs Arrokoth and Kleopatra and to comet 103P/Hartley, and the results showed that the GDSM outperforms the point-mass modeling and the dipole-segment modeling. The GDSM provides more accuracy while preserving the advantage of low computational cost. It becomes a powerful and efficient tool for studying the gravitational environment around small celestial bodies and takes a step toward a more realistic scenario while still allowing for a computationally unburdened code in comparison with the expensive polyhedral model. This can be useful for early stages of a mission design for a visit to this type of asteroid. Furthermore, the GDSM can also be adopted in particular for real-time orbital corrections. This needs to be evaluated onboard with the few computational resources.

From the GDSM, we calculated the initial conditions and integrated the heteroclinic orbits that connect the unstable triangular points for Arrokoth, Kleopatra, and comet 103P/Hartley. For each case, four heteroclinic orbits were found, two stable and two unstable orbits. The orbits with smaller amplitudes (Arrokoth and 103P/Hartley) are around bodies whose distance from the origin to the extremity of the body is greater, and smaller amplitudes for the opposite. For Kleopatra, whose distance from the extremities is more balanced, the orbits tend to have balanced amplitudes as well. The combination of each pair of heteroclinic orbits forms a symmetrical planar periodic orbit around the body.

\begin{acknowledgements}
The authors wish to express their appreciation for the support provided by grant \#309089/2021-2, \#443116/2023-7, and \#201718/2025-1 from the National Council for Scientific and Technological Development (CNPq), grant \#2022/11783-5 from São Paulo Research Foundation (FAPESP).
A.~Amarante and A.~Ferreira thank the financial support of the São Paulo Research Foundation (FAPESP) [grants \#2023/11781-5 \& \#2025/15438-9]. Additionally, Santos, L. B. T. would like to thank the support of the Polytechnic School of the University of Pernambuco (POLI-UPE) and the PostGrad Program in Systems Engineering (PPGES).
The authors also acknowledge the financial support of the Coordination for the Improvement of Higher Education Personnel (CAPES) - Brazil (finance code 001). This research was also supported by computational resources supplied by the Center for Scientific Computing (NCC/GridUNESP) of the São Paulo State University (UNESP) and the Center for Mathematical Sciences Applied to Industry (CeMEAI), funded by FAPESP [grant \#2013/07375-0].
F.M. thanks the fellowship granted by FAPESP (\#2024/16260-6).
E. Tresaco thanks Grant PID2024-156002NB-I00 funded by MICIU/AEI/10.13039/501100011033/FEDER, UE.
\end{acknowledgements}

% WARNING
%-------------------------------------------------------------------
% Please note that we have included the references to the file aa.dem in
% order to compile it, but we ask you to:
%
% - use BibTeX with the regular commands:
  \bibliographystyle{aa} % style aa.bst
  \bibliography{ref_leo} % your references Yourfile.bib

\end{document}